\documentclass[
	12pt,				
    openany,  
    twoside,			
	a4paper,			
	english,			
	french,				
	spanish,			
	brazil,				
	]{abntex2}

\usepackage[top=3cm, bottom=2cm, left=3cm, right=2cm]{geometry} 
\usepackage{lmodern}			
\usepackage[T1]{fontenc}		
\usepackage{calligra}
\usepackage[utf8]{inputenc}		
\usepackage{indentfirst}		
\usepackage{color}				
\usepackage{graphicx}			
\usepackage{microtype} 			
\usepackage{amssymb, amsmath, pxfonts} 
\usepackage{mathrsfs} 
\usepackage[normalem]{ulem} 
\usepackage{mathrsfs} 
\usepackage{graphicx} 
\usepackage[footnotesize]{caption}
\usepackage{setspace} 
\usepackage{cite}
\usepackage{verbatim}
\usepackage{calrsfs}
\usepackage{mathtools}
\usepackage{cleveref}
\usepackage{ragged2e}
\usepackage{scrextend}

\makeatletter
\let\@currsize\normalsize
\makeatother

\usepackage{lipsum}				

\definecolor{blue}{RGB}{41,5,195}

\makeatletter
\hypersetup{
		pdftitle={\@title}, 
		pdfauthor={\@author},
    	pdfsubject={\imprimirpreambulo},
	    pdfcreator={LaTeX with abnTeX2},
		pdfkeywords={abnt}{latex}{abntex}{abntex2}{projeto de pesquisa}, 
		colorlinks=true,       		
    	linkcolor=blue,          	
    	citecolor=blue,        		
    	filecolor=magenta,      		
		urlcolor=blue,
		bookmarksdepth=4
}
\makeatother

\makeindex
\begin{document}

\selectlanguage{brazil}

\frenchspacing

\pagebreak

\begin{center}
\begin{Large}
\textbf{PRÁTICA DE FÍSICA EXPERIMENTAL \\}
\end{Large}
\vspace{5cm}
\begin{Large}
\textbf{Aplicação do Método Científico no Estudo \\
da Pandemia de COVID-19 \\}
\end{Large}
\end{center}
Colégio Tiradentes da Polícia Militar - CTPM / São João del Rei - MG\\
Prof. Dr. Raul Guilherme Batista Mendes
\vspace{5cm}
\pagebreak
  
 \section*{Método científico; germinação e crescimento do feijão} \label{MetCientífico}

\underline{Introdução}

Durante o período histórico conhecido como Renascimento, iniciado na região da atual Itália e que se estendeu, aproximadamente, entre os anos de 1400 e 1600, a representação por pinturas em tela passou por uma de suas maiores evoluções artística. Renomados pintores como, Donatello (1346-1466), Piero della Francesca (1410-1492), Andrea Mantegna (1431-1506), Michelangelo Buonarroti (1475-1564) e Rafael (1483-1520) foram os principais responsáveis pelo desenvolvimento do conceito atualmente conhecido como \textit{perspectiva}. Adotando-se este novo conceito de pintura, a descrição de um determinado cenário era feita conservando-se a proporcionalidade dos tamanhos dos objetos. A tela de pintura passaram a evidenciar a profundidade dos corpos e ambientes, possibilitando compreensões trí-dimensionais da imagem (altura, largura e profundidade) \cite{positivo, atlas}.

\begin{figure}[ht]
\centering
  \includegraphics[scale=0.5]{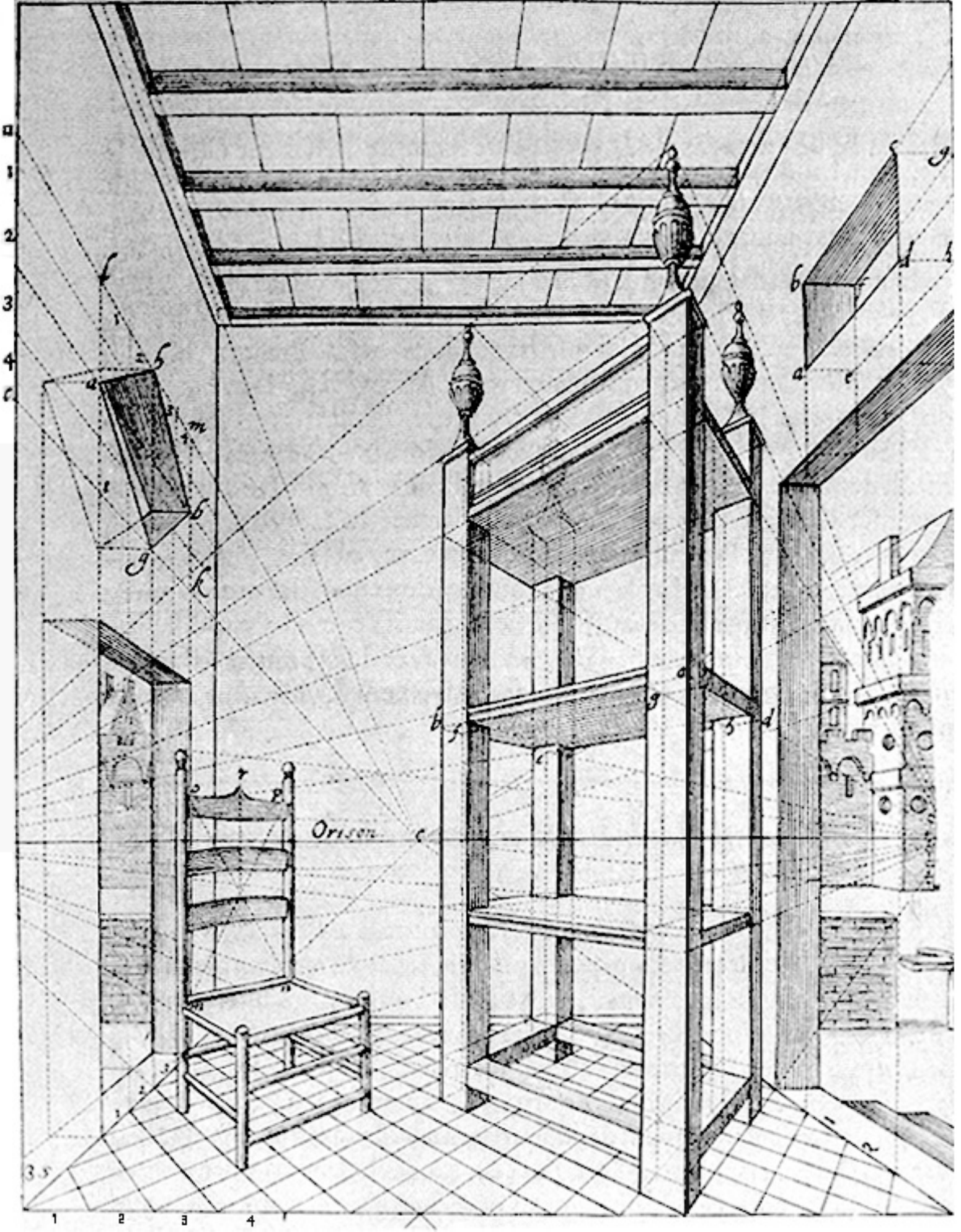}
  \captionof{figure}{Exemplo de obra de arte feita em tela explorando a \textit{perspectiva} \cite{perspectiva}. Os desenhos são feitos respeitando suas dimensões de escala, ou seja, mantendo seus tamanhos proporcionais, o que garante melhor leitura de profundidade na imagem.}
  \label{percpectiva}
\end{figure}

Este aprimoramento da pintura na descrição de cenários, pessoas e objetos também estendeu-se a outras áreas, além das artes; estendeu-se às ciências. Às ciências biológicas foram implementadas melhores técnicas de descrição nos desenho da anatomia humana e animal. Nas ciências exatas, foi a partir deste ponto que os experimentos de investigação da natureza ganharam melhor descrição experimental, principalmente, através de desenhos ilustrativos da forma como as práticas eram realizadas. Os principais filósofos/cientistas desta época a se utilizarem desta técnica descritiva do experimento foram Galileu Galilei (1564-1642), René Descartes (1596-1650), Giordano Bruno (1548-1600) e Francis Bacon (1561-1626). Estes, principalmente os dois primeiros, foram os grandes responsáveis pela criação o conceito conhecido atualmente como \textit{Método Científico}.

Com o método científico pretende-se padronizar a prática experimental dentro de certos preceitos, a fim de se garantir que as observações de um determinado experimento tenham validade para a definição, ou constatação, de uma teoria científica ou hipótese. Além disso, o método científico possibilita a reprodução de uma prática experimental, seja pelo mesmo pesquisador ou por outro, assegurando uma confiabilidade mínima para a comparação dos dados obtidos em diferentes instantes e locais.

Nesta prática experimental o aluno deverá acompanhar a germinação e o crescimento de grãos de feijão em algodão umedecido. Submetendo os grãos a diferentes condições (diferentes variáveis), o aluno deverá constatar o processo de causa e efeito sobre o crescimento do feijão.  

\underline{Tópicos a serem abordados}

Método científico para a realização de experimentos, correlação entre dados experimentais, curva de distribuição estatística.

\underline{Material utilizado} \\
- grãos de feijão \\
- algodão \\
- copos descartáveis \\
- colher de sopa e \\
- água 

\underline{Execução da prática}

Em sua casa o aluno deverá fazer a germinação de 10 grãos de feijão. Estes 10 grãos deverão ser divididos em dois grupos de 5 grãos de feijão cada, o grupo A e o grupo B. Todos os 10 grão de feijão deverão ser plantados separadamente, porém os 5 grãos do grupo A deverão ser plantados mantendo-se um mesmo padrão, ou seja, plantados de forma mais semelhante possível. Para isto, o aluno deverá replicar nos feijões do grupo A a técnica de plantio de feijão em algodão descrita no vídeo disponibilizados na plataforma Youtube: \href{https://www.youtube.com/watch?v=cyhPqh8g4zk}{https://www.youtube.com/watch?v=cyhPqh8g4zk}. Neste vídeo, a técnica de plantio de feijão em algodão é descrita nos primeiros minutos da gravação e a etapa de replantio em um vaso com terra não será utilizada na prática experimental que estamos desenvolvendo agora. Caso o aluno não tenha copos plásticos descartáveis à sua disposição, poderão ser utilizados copos de vidro, desde que se dê preferência para que os copos usados no grupo A sejam todos iguais. Todos os cinco feijões do grupo A, cada um plantado em um pedaço de algodão dentro de um copo, deveram ser colocados próximo a uma mesma janela para que recebam a luz do Sol pela manhã. A quantidade de algodão usado para cada feijão do grupo A deve ser a mesma e sempre que estes algodões estiverem secos deve-se umedecê-los com a mesma quantidade de água, uma colher de sopa e água todas as manhãs. Mantenha os algodões umedecidos, porém sem excesso de água. \textbf{A ideia é que todos os alunos envolvidos no experimento germinem seus feijões do grupo A de forma mais parecida possível. Por isso, torna-se tão importante que todos os alunos tentem replicar ao máximo os procedimentos exibidos no vídeo (tamanho e cor do copo plástico, quantidade de algodão, umedecer completamente o algodão antes de colocar o feijão e depois manter úmido adicionando uma colher de sopa de água por dia pela manhã, ...).} A única diferença que se pede dos passos adotados no vídeo é que sejam utilizados, preferencialmente, feijões marrons (feijões carioquinha) ou, se não, feijões vermelhos ou, se não, então, feijões pretos.

Quanto aos grãos de feijão do grupo B, cada aluno deverá plantar seus 5 feijões de forma aleatória. Obviamente, deve-se adotar procedimentos que possibilitem a germinação de cada grão de feijão, não exceder na água e no calor absorvido, por exemplo. Porém, a quantidade e a frequência da umidificação do algodão, a intensidade de luz e período de exposição direta, o tipo feijão, a quantidade de grãos de feijão por pedaço de algodão, entre outras variáveis, devem ser definidas para cada grão de feijão pelo próprio aluno. A ideia para os grãos do grupo B é que eles germinem, mas que todas as germinações realizadas pelos alunos tenham a menor semelhanças possível (baixa correlação entre as germinações). 

O aluno deverá acompanhar os processos de germinação e observar em qual dia, a partir do dia de plantio, cada um de seus 10 grãos de feijão apresentou sua primeira folha. Para a coleta dos dados o aluno deverá preencher as tabelas apresentadas abaixo, na seção de \textit{Coleta de dados}.

\underline{Coleta dos dados}

Procure preencher as informações abaixo de forma clara e direta. Lembre-se, um dos objetivos do método científico é descrever seu experimento tão bem, ao ponto que possibilite que outra pessoa o repita posteriormente de forma mais fidedigna possível.

\textbf{A duração da coleta de dados deve se estender por, pelo menos, 15 dias, para que se observe o maior número de germinações possíveis!} 
\\
\\
\textbf{Grupo A} \\ \\
cor dos feijões: \\
\_\_\_\_\_\_\_\_\_\_\_\_\_\_\_\_\_\_\_\_\_\_\_\_\_\_\_\_\_\_\_\_\_\_\_\_\_\_\_\_\_\_\_\_\_\_\_\_\_\_\_\_\_\_\_\_\_\_\_\_\_\_\_\_\_\_\_\_\_\_\_\_\_\_\_\_\_\_\_\_ \\ \\
tipo, cor e volume dos copos usados (quantos ml cabe em cada copo): \\
\_\_\_\_\_\_\_\_\_\_\_\_\_\_\_\_\_\_\_\_\_\_\_\_\_\_\_\_\_\_\_\_\_\_\_\_\_\_\_\_\_\_\_\_\_\_\_\_\_\_\_\_\_\_\_\_\_\_\_\_\_\_\_\_\_\_\_\_\_\_\_\_\_\_\_\_\_\_\_\_ \\ 
\_\_\_\_\_\_\_\_\_\_\_\_\_\_\_\_\_\_\_\_\_\_\_\_\_\_\_\_\_\_\_\_\_\_\_\_\_\_\_\_\_\_\_\_\_\_\_\_\_\_\_\_\_\_\_\_\_\_\_\_\_\_\_\_\_\_\_\_\_\_\_\_\_\_\_\_\_\_\_\_ \\ \\ \\
quantidade de água usada e frequência da umidificação: \\
\_\_\_\_\_\_\_\_\_\_\_\_\_\_\_\_\_\_\_\_\_\_\_\_\_\_\_\_\_\_\_\_\_\_\_\_\_\_\_\_\_\_\_\_\_\_\_\_\_\_\_\_\_\_\_\_\_\_\_\_\_\_\_\_\_\_\_\_\_\_\_\_\_\_\_\_\_\_\_\_ \\
\_\_\_\_\_\_\_\_\_\_\_\_\_\_\_\_\_\_\_\_\_\_\_\_\_\_\_\_\_\_\_\_\_\_\_\_\_\_\_\_\_\_\_\_\_\_\_\_\_\_\_\_\_\_\_\_\_\_\_\_\_\_\_\_\_\_\_\_\_\_\_\_\_\_\_\_\_\_\_\_ \\
luminosidade sobre os grãos (forte, moderada ou fraca): \\
\_\_\_\_\_\_\_\_\_\_\_\_\_\_\_\_\_\_\_\_\_\_\_\_\_\_\_\_\_\_\_\_\_\_\_\_\_\_\_\_\_\_\_\_\_\_\_\_\_\_\_\_\_\_\_\_\_\_\_\_\_\_\_\_\_\_\_\_\_\_\_\_\_\_\_\_\_\_\_\_ \\
\_\_\_\_\_\_\_\_\_\_\_\_\_\_\_\_\_\_\_\_\_\_\_\_\_\_\_\_\_\_\_\_\_\_\_\_\_\_\_\_\_\_\_\_\_\_\_\_\_\_\_\_\_\_\_\_\_\_\_\_\_\_\_\_\_\_\_\_\_\_\_\_\_\_\_\_\_\_\_\_ \\

\begin{table}[ht]
\begin{tabular}{|p{1cm}|p{4cm}|}
\hline
feijão &dia em que a primeira folha apareceu\\
\hline
1 & \\
\hline
2 & \\
\hline
3 & \\
\hline
4 & \\
\hline
5 & \\
\hline
\end{tabular}
\end{table}
 \vspace{1,4cm}

\textbf{Grupo B}

\textbf{Feijão 1} \\ \\
dia em que a primeira folha apareceu: \\
\_\_\_\_\_\_\_\_\_\_\_\_\_\_\_\_\_\_\_\_\_\_\_\_\_\_\_\_\_\_\_\_\_\_\_\_\_\_\_\_\_\_\_\_\_\_\_\_\_\_\_\_\_\_\_\_\_\_\_\_\_\_\_\_\_\_\_\_\_\_\_\_\_\_\_\_\_\_\_\_ \\
cor do feijão: \\
\_\_\_\_\_\_\_\_\_\_\_\_\_\_\_\_\_\_\_\_\_\_\_\_\_\_\_\_\_\_\_\_\_\_\_\_\_\_\_\_\_\_\_\_\_\_\_\_\_\_\_\_\_\_\_\_\_\_\_\_\_\_\_\_\_\_\_\_\_\_\_\_\_\_\_\_\_\_\_\_ \\\
tipo, cor e volume do copo usado: \\
\_\_\_\_\_\_\_\_\_\_\_\_\_\_\_\_\_\_\_\_\_\_\_\_\_\_\_\_\_\_\_\_\_\_\_\_\_\_\_\_\_\_\_\_\_\_\_\_\_\_\_\_\_\_\_\_\_\_\_\_\_\_\_\_\_\_\_\_\_\_\_\_\_\_\_\_\_\_\_\_ \\ 
\_\_\_\_\_\_\_\_\_\_\_\_\_\_\_\_\_\_\_\_\_\_\_\_\_\_\_\_\_\_\_\_\_\_\_\_\_\_\_\_\_\_\_\_\_\_\_\_\_\_\_\_\_\_\_\_\_\_\_\_\_\_\_\_\_\_\_\_\_\_\_\_\_\_\_\_\_\_\_\_ \\ \\
quantidade de água usada e frequência da umidificação: \\
\_\_\_\_\_\_\_\_\_\_\_\_\_\_\_\_\_\_\_\_\_\_\_\_\_\_\_\_\_\_\_\_\_\_\_\_\_\_\_\_\_\_\_\_\_\_\_\_\_\_\_\_\_\_\_\_\_\_\_\_\_\_\_\_\_\_\_\_\_\_\_\_\_\_\_\_\_\_\_\_ \\
\_\_\_\_\_\_\_\_\_\_\_\_\_\_\_\_\_\_\_\_\_\_\_\_\_\_\_\_\_\_\_\_\_\_\_\_\_\_\_\_\_\_\_\_\_\_\_\_\_\_\_\_\_\_\_\_\_\_\_\_\_\_\_\_\_\_\_\_\_\_\_\_\_\_\_\_\_\_\_\_ \\
luminosidade: \\
\_\_\_\_\_\_\_\_\_\_\_\_\_\_\_\_\_\_\_\_\_\_\_\_\_\_\_\_\_\_\_\_\_\_\_\_\_\_\_\_\_\_\_\_\_\_\_\_\_\_\_\_\_\_\_\_\_\_\_\_\_\_\_\_\_\_\_\_\_\_\_\_\_\_\_\_\_\_\_\_ \\
\_\_\_\_\_\_\_\_\_\_\_\_\_\_\_\_\_\_\_\_\_\_\_\_\_\_\_\_\_\_\_\_\_\_\_\_\_\_\_\_\_\_\_\_\_\_\_\_\_\_\_\_\_\_\_\_\_\_\_\_\_\_\_\_\_\_\_\_\_\_\_\_\_\_\_\_\_\_\_\_ \\

\textbf{Feijão 2} \\ \\
dia em que a primeira folha apareceu: \\
\_\_\_\_\_\_\_\_\_\_\_\_\_\_\_\_\_\_\_\_\_\_\_\_\_\_\_\_\_\_\_\_\_\_\_\_\_\_\_\_\_\_\_\_\_\_\_\_\_\_\_\_\_\_\_\_\_\_\_\_\_\_\_\_\_\_\_\_\_\_\_\_\_\_\_\_\_\_\_\_ \\
cor do feijão: \\
\_\_\_\_\_\_\_\_\_\_\_\_\_\_\_\_\_\_\_\_\_\_\_\_\_\_\_\_\_\_\_\_\_\_\_\_\_\_\_\_\_\_\_\_\_\_\_\_\_\_\_\_\_\_\_\_\_\_\_\_\_\_\_\_\_\_\_\_\_\_\_\_\_\_\_\_\_\_\_\_ \\\
tipo, cor e volume do copo usado: \\
\_\_\_\_\_\_\_\_\_\_\_\_\_\_\_\_\_\_\_\_\_\_\_\_\_\_\_\_\_\_\_\_\_\_\_\_\_\_\_\_\_\_\_\_\_\_\_\_\_\_\_\_\_\_\_\_\_\_\_\_\_\_\_\_\_\_\_\_\_\_\_\_\_\_\_\_\_\_\_\_ \\ 
\_\_\_\_\_\_\_\_\_\_\_\_\_\_\_\_\_\_\_\_\_\_\_\_\_\_\_\_\_\_\_\_\_\_\_\_\_\_\_\_\_\_\_\_\_\_\_\_\_\_\_\_\_\_\_\_\_\_\_\_\_\_\_\_\_\_\_\_\_\_\_\_\_\_\_\_\_\_\_\_ \\ \\
quantidade de água usada e frequência da umidificação: \\
\_\_\_\_\_\_\_\_\_\_\_\_\_\_\_\_\_\_\_\_\_\_\_\_\_\_\_\_\_\_\_\_\_\_\_\_\_\_\_\_\_\_\_\_\_\_\_\_\_\_\_\_\_\_\_\_\_\_\_\_\_\_\_\_\_\_\_\_\_\_\_\_\_\_\_\_\_\_\_\_ \\
\_\_\_\_\_\_\_\_\_\_\_\_\_\_\_\_\_\_\_\_\_\_\_\_\_\_\_\_\_\_\_\_\_\_\_\_\_\_\_\_\_\_\_\_\_\_\_\_\_\_\_\_\_\_\_\_\_\_\_\_\_\_\_\_\_\_\_\_\_\_\_\_\_\_\_\_\_\_\_\_ \\
luminosidade: \\
\_\_\_\_\_\_\_\_\_\_\_\_\_\_\_\_\_\_\_\_\_\_\_\_\_\_\_\_\_\_\_\_\_\_\_\_\_\_\_\_\_\_\_\_\_\_\_\_\_\_\_\_\_\_\_\_\_\_\_\_\_\_\_\_\_\_\_\_\_\_\_\_\_\_\_\_\_\_\_\_ \\
\_\_\_\_\_\_\_\_\_\_\_\_\_\_\_\_\_\_\_\_\_\_\_\_\_\_\_\_\_\_\_\_\_\_\_\_\_\_\_\_\_\_\_\_\_\_\_\_\_\_\_\_\_\_\_\_\_\_\_\_\_\_\_\_\_\_\_\_\_\_\_\_\_\_\_\_\_\_\_\_ \\

\textbf{Feijão 3} \\ \\
dia em que a primeira folha apareceu: \\
\_\_\_\_\_\_\_\_\_\_\_\_\_\_\_\_\_\_\_\_\_\_\_\_\_\_\_\_\_\_\_\_\_\_\_\_\_\_\_\_\_\_\_\_\_\_\_\_\_\_\_\_\_\_\_\_\_\_\_\_\_\_\_\_\_\_\_\_\_\_\_\_\_\_\_\_\_\_\_\_ \\
cor do feijão: \\
\_\_\_\_\_\_\_\_\_\_\_\_\_\_\_\_\_\_\_\_\_\_\_\_\_\_\_\_\_\_\_\_\_\_\_\_\_\_\_\_\_\_\_\_\_\_\_\_\_\_\_\_\_\_\_\_\_\_\_\_\_\_\_\_\_\_\_\_\_\_\_\_\_\_\_\_\_\_\_\_ \\\
tipo, cor e volume do copo usado: \\
\_\_\_\_\_\_\_\_\_\_\_\_\_\_\_\_\_\_\_\_\_\_\_\_\_\_\_\_\_\_\_\_\_\_\_\_\_\_\_\_\_\_\_\_\_\_\_\_\_\_\_\_\_\_\_\_\_\_\_\_\_\_\_\_\_\_\_\_\_\_\_\_\_\_\_\_\_\_\_\_ \\ 
\_\_\_\_\_\_\_\_\_\_\_\_\_\_\_\_\_\_\_\_\_\_\_\_\_\_\_\_\_\_\_\_\_\_\_\_\_\_\_\_\_\_\_\_\_\_\_\_\_\_\_\_\_\_\_\_\_\_\_\_\_\_\_\_\_\_\_\_\_\_\_\_\_\_\_\_\_\_\_\_ \\ \\
quantidade de água usada e frequência da umidificação: \\
\_\_\_\_\_\_\_\_\_\_\_\_\_\_\_\_\_\_\_\_\_\_\_\_\_\_\_\_\_\_\_\_\_\_\_\_\_\_\_\_\_\_\_\_\_\_\_\_\_\_\_\_\_\_\_\_\_\_\_\_\_\_\_\_\_\_\_\_\_\_\_\_\_\_\_\_\_\_\_\_ \\
\_\_\_\_\_\_\_\_\_\_\_\_\_\_\_\_\_\_\_\_\_\_\_\_\_\_\_\_\_\_\_\_\_\_\_\_\_\_\_\_\_\_\_\_\_\_\_\_\_\_\_\_\_\_\_\_\_\_\_\_\_\_\_\_\_\_\_\_\_\_\_\_\_\_\_\_\_\_\_\_ \\
luminosidade: \\
\_\_\_\_\_\_\_\_\_\_\_\_\_\_\_\_\_\_\_\_\_\_\_\_\_\_\_\_\_\_\_\_\_\_\_\_\_\_\_\_\_\_\_\_\_\_\_\_\_\_\_\_\_\_\_\_\_\_\_\_\_\_\_\_\_\_\_\_\_\_\_\_\_\_\_\_\_\_\_\_ \\
\_\_\_\_\_\_\_\_\_\_\_\_\_\_\_\_\_\_\_\_\_\_\_\_\_\_\_\_\_\_\_\_\_\_\_\_\_\_\_\_\_\_\_\_\_\_\_\_\_\_\_\_\_\_\_\_\_\_\_\_\_\_\_\_\_\_\_\_\_\_\_\_\_\_\_\_\_\_\_\_ \\

\textbf{Feijão 4} \\ \\
dia em que a primeira folha apareceu: \\
\_\_\_\_\_\_\_\_\_\_\_\_\_\_\_\_\_\_\_\_\_\_\_\_\_\_\_\_\_\_\_\_\_\_\_\_\_\_\_\_\_\_\_\_\_\_\_\_\_\_\_\_\_\_\_\_\_\_\_\_\_\_\_\_\_\_\_\_\_\_\_\_\_\_\_\_\_\_\_\_ \\
cor do feijão: \\
\_\_\_\_\_\_\_\_\_\_\_\_\_\_\_\_\_\_\_\_\_\_\_\_\_\_\_\_\_\_\_\_\_\_\_\_\_\_\_\_\_\_\_\_\_\_\_\_\_\_\_\_\_\_\_\_\_\_\_\_\_\_\_\_\_\_\_\_\_\_\_\_\_\_\_\_\_\_\_\_ \\\
tipo, cor e volume do copo usado: \\
\_\_\_\_\_\_\_\_\_\_\_\_\_\_\_\_\_\_\_\_\_\_\_\_\_\_\_\_\_\_\_\_\_\_\_\_\_\_\_\_\_\_\_\_\_\_\_\_\_\_\_\_\_\_\_\_\_\_\_\_\_\_\_\_\_\_\_\_\_\_\_\_\_\_\_\_\_\_\_\_ \\ 
\_\_\_\_\_\_\_\_\_\_\_\_\_\_\_\_\_\_\_\_\_\_\_\_\_\_\_\_\_\_\_\_\_\_\_\_\_\_\_\_\_\_\_\_\_\_\_\_\_\_\_\_\_\_\_\_\_\_\_\_\_\_\_\_\_\_\_\_\_\_\_\_\_\_\_\_\_\_\_\_ \\ \\
quantidade de água usada e frequência da umidificação: \\
\_\_\_\_\_\_\_\_\_\_\_\_\_\_\_\_\_\_\_\_\_\_\_\_\_\_\_\_\_\_\_\_\_\_\_\_\_\_\_\_\_\_\_\_\_\_\_\_\_\_\_\_\_\_\_\_\_\_\_\_\_\_\_\_\_\_\_\_\_\_\_\_\_\_\_\_\_\_\_\_ \\
\_\_\_\_\_\_\_\_\_\_\_\_\_\_\_\_\_\_\_\_\_\_\_\_\_\_\_\_\_\_\_\_\_\_\_\_\_\_\_\_\_\_\_\_\_\_\_\_\_\_\_\_\_\_\_\_\_\_\_\_\_\_\_\_\_\_\_\_\_\_\_\_\_\_\_\_\_\_\_\_ \\
luminosidade: \\
\_\_\_\_\_\_\_\_\_\_\_\_\_\_\_\_\_\_\_\_\_\_\_\_\_\_\_\_\_\_\_\_\_\_\_\_\_\_\_\_\_\_\_\_\_\_\_\_\_\_\_\_\_\_\_\_\_\_\_\_\_\_\_\_\_\_\_\_\_\_\_\_\_\_\_\_\_\_\_\_ \\
\_\_\_\_\_\_\_\_\_\_\_\_\_\_\_\_\_\_\_\_\_\_\_\_\_\_\_\_\_\_\_\_\_\_\_\_\_\_\_\_\_\_\_\_\_\_\_\_\_\_\_\_\_\_\_\_\_\_\_\_\_\_\_\_\_\_\_\_\_\_\_\_\_\_\_\_\_\_\_\_ \\

\textbf{Feijão 5} \\ \\
dia em que a primeira folha apareceu: \\
\_\_\_\_\_\_\_\_\_\_\_\_\_\_\_\_\_\_\_\_\_\_\_\_\_\_\_\_\_\_\_\_\_\_\_\_\_\_\_\_\_\_\_\_\_\_\_\_\_\_\_\_\_\_\_\_\_\_\_\_\_\_\_\_\_\_\_\_\_\_\_\_\_\_\_\_\_\_\_\_ \\
cor dos feijão: \\
\_\_\_\_\_\_\_\_\_\_\_\_\_\_\_\_\_\_\_\_\_\_\_\_\_\_\_\_\_\_\_\_\_\_\_\_\_\_\_\_\_\_\_\_\_\_\_\_\_\_\_\_\_\_\_\_\_\_\_\_\_\_\_\_\_\_\_\_\_\_\_\_\_\_\_\_\_\_\_\_ \\\
tipo, cor e volume do copo usado: \\
\_\_\_\_\_\_\_\_\_\_\_\_\_\_\_\_\_\_\_\_\_\_\_\_\_\_\_\_\_\_\_\_\_\_\_\_\_\_\_\_\_\_\_\_\_\_\_\_\_\_\_\_\_\_\_\_\_\_\_\_\_\_\_\_\_\_\_\_\_\_\_\_\_\_\_\_\_\_\_\_ \\ 
\_\_\_\_\_\_\_\_\_\_\_\_\_\_\_\_\_\_\_\_\_\_\_\_\_\_\_\_\_\_\_\_\_\_\_\_\_\_\_\_\_\_\_\_\_\_\_\_\_\_\_\_\_\_\_\_\_\_\_\_\_\_\_\_\_\_\_\_\_\_\_\_\_\_\_\_\_\_\_\_ \\ \\
quantidade de água usada e frequência da umidificação: \\
\_\_\_\_\_\_\_\_\_\_\_\_\_\_\_\_\_\_\_\_\_\_\_\_\_\_\_\_\_\_\_\_\_\_\_\_\_\_\_\_\_\_\_\_\_\_\_\_\_\_\_\_\_\_\_\_\_\_\_\_\_\_\_\_\_\_\_\_\_\_\_\_\_\_\_\_\_\_\_\_ \\
\_\_\_\_\_\_\_\_\_\_\_\_\_\_\_\_\_\_\_\_\_\_\_\_\_\_\_\_\_\_\_\_\_\_\_\_\_\_\_\_\_\_\_\_\_\_\_\_\_\_\_\_\_\_\_\_\_\_\_\_\_\_\_\_\_\_\_\_\_\_\_\_\_\_\_\_\_\_\_\_ \\
luminosidade: \\
\_\_\_\_\_\_\_\_\_\_\_\_\_\_\_\_\_\_\_\_\_\_\_\_\_\_\_\_\_\_\_\_\_\_\_\_\_\_\_\_\_\_\_\_\_\_\_\_\_\_\_\_\_\_\_\_\_\_\_\_\_\_\_\_\_\_\_\_\_\_\_\_\_\_\_\_\_\_\_\_ \\
\_\_\_\_\_\_\_\_\_\_\_\_\_\_\_\_\_\_\_\_\_\_\_\_\_\_\_\_\_\_\_\_\_\_\_\_\_\_\_\_\_\_\_\_\_\_\_\_\_\_\_\_\_\_\_\_\_\_\_\_\_\_\_\_\_\_\_\_\_\_\_\_\_\_\_\_\_\_\_\_ \\

\underline{Expectativa quanto aos dados}

Visto que os grãos de feijão do grupo A de todos os alunos germinaram em condições muito próximas, é de se esperar que os dias gastos para a aparição da primeira folha de cada um destes grãos tenham uma relação grande entre si. Desta forma, devemos ter uma maior concentração destes eventos dentro de um intervalo de tempo mais estreito. Quanto aos tempos de aparições das primeiras folhas dos grãos do grupo B, espera-se uma baixa relação entre si. Assim, tais eventos devem se mostrar mais dispersas ao longo do período de 15 dias.

\underline{Discussão dos dados e conclusão}

Coordenados pelo professor, os alunos devem unificar todos os tempos observados em dois grupos, formando os grupos $A_{total}$ e $B_{total}$. Os gráficos \textit{porcentagens de aparição da 1ª folha} x \textit{Nº de dias} deverão ser construídos para cada um destes dois grupos.

Conforme descrito anteriormente, nos grãos do grupo A tentou-se garantir, ao máximo, que todas as germinações ocorressem em condições mais próximas possíveis, ou seja, que ocorressem com alta correlação entre si. Ao passo que, nas germinações do grupo B não se garantiu correlação nenhuma entre os eventos mensurados.

\begin{figure}[ht]
\centering
  \includegraphics[scale=0.15]{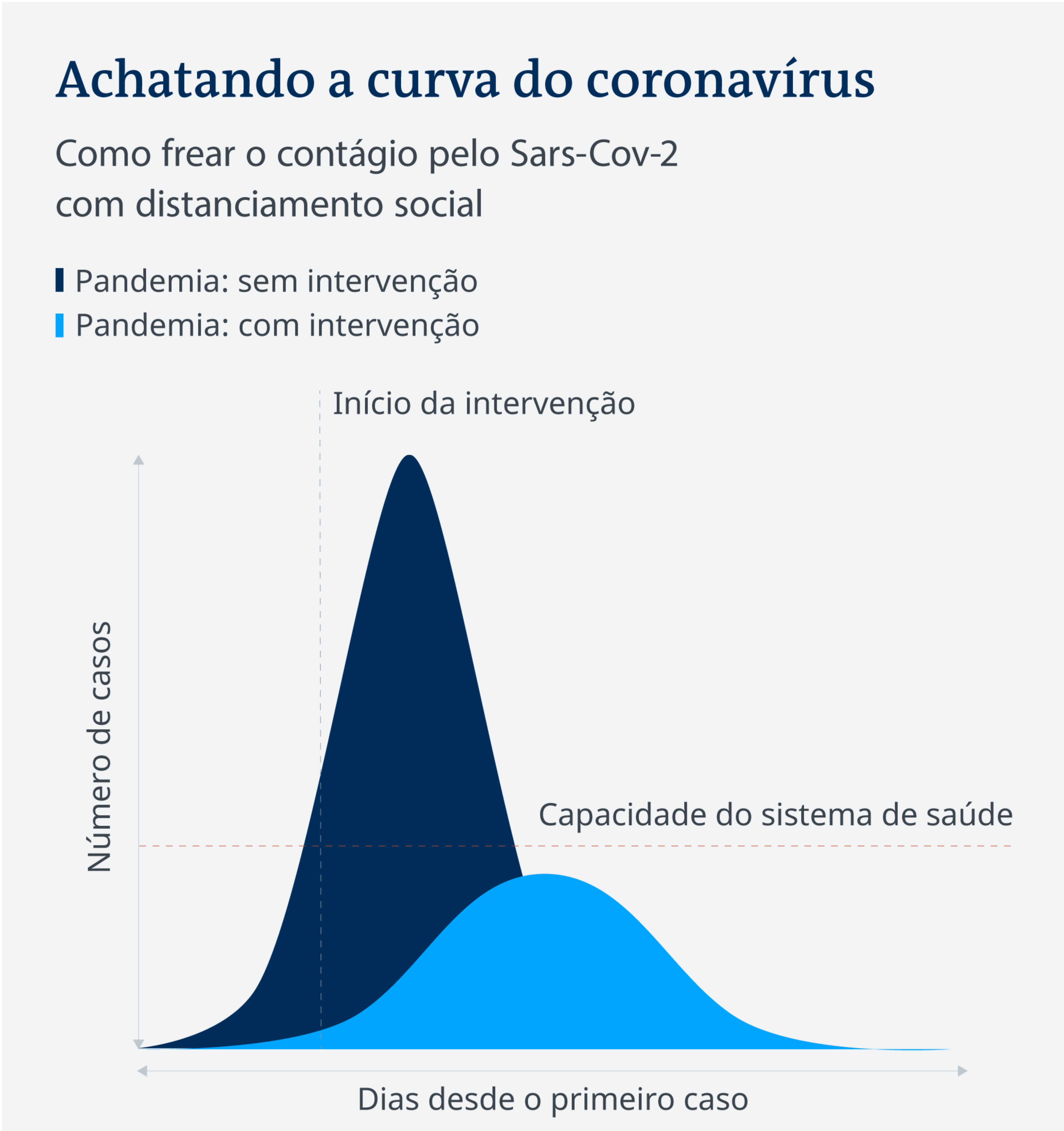}
  \captionof{figure}{Fonte: Centro de Controle e Prevenção de Doenças (EUA), Instituto Dalle Lana de Saúde Pública. Fonte: \href{https://www.dw.com/pt-br/os-n\%C3\%BAmeros-sobre-a-pandemia-de-coronav\%C3\%ADrus/a-52848559}{https://www.dw.com/pt-br/os-n\%C3\%BAmeros-sobre-a-pandemia-de-coronav\%C3\%ADrus/a-52848559}).}
  \label{curva_corona}
\end{figure}

Tomemos como comparação a conjuntura atual, com a pandemia causada pelo vírus Sars-Cov-2, da família coronavírus e responsável pela doença Covid-19. Supondo que todos os indivíduos de uma determinada população em algum momento ao longo do ano de 2020 serão infectados pelo Sars-Cov-2, então a Fig-\ref{curva_corona} nos mostra duas situações distintas para a \textit{quantidade de infectados} (Número de casos) em função do \textit{tempo gasto para se infectar} (Dias desde o primeiro caso). Como podemos relacionar as formas das curvas experimentais obtidas para os grupos $A_{total}$ e $B_{total}$ com as formas das curvas da pandemia ``sem intervenção'' e ``com intervenção'', apresentadas na Fig-\ref{curva_corona}?

Se definirmos como eventos probabilísticos o \textit{ato de brotar uma folha de feijão} e o \textit{ato de se infectar pelo Sars-Cov-2}, baseando-se na discussão feita a cerca dos resultados experimentais que obtivemos das germinações, em qual das duas curvas da pandemia apresentadas na Fig-\ref{curva_corona} observa-se maior correlação entre os atos de se infectar pelo Sars-Cov-2 em uma população? Quais as características das curvas que evidenciam isto?

Qual a conclusão que se pode tirar deste experimento, sobre:

a) A importância do método científico para a interpretação de dados científicos?

b) A importância de se poder fazer um trabalho experimental em grupo, buscando-se maior quantidade de observações a cerca de uma questão a ser investigada?

\clearpage

\bibliographystyle{agsm}

\end{document}